\documentclass[10pt,conference]{IEEEtran}
\IEEEoverridecommandlockouts
\usepackage{amsmath,amssymb,amsfonts}
\usepackage{algorithmic}
\usepackage{graphicx}
\usepackage{xcolor}
\usepackage{siunitx}
\usepackage[capitalize]{cleveref}
\usepackage{booktabs}
\usepackage{tabularx}
\usepackage{pgfplots}
\usepackage{microtype}
\usepackage{fancyhdr}
\usepackage[style=ieee, maxcitenames=2, mincitenames=1, citestyle=numeric-comp,backend=bibtex]{biblatex}
\fancyfootoffset{0pt}
\fancypagestyle{firstpage}{
    \fancyhf{}
    \fancyfoot[L]{\footnotesize © 2026 IEEE.  Personal use of this material is permitted.  Permission from IEEE must be obtained for all other uses, in any current or future media, including reprinting/republishing this material for advertising or promotional purposes, creating new collective works, for resale or redistribution to servers or lists, or reuse of any copyrighted component of this work in other works.}
    
}

\AtEveryBibitem{
 	\clearfield{url}
 	\clearfield{doi}
\ifentrytype{inproceedings}{
 	\clearlist{address}
 	\clearlist{publisher}
 	\clearname{editor}
 	\clearlist{organization}
 	\clearfield{pages}
 	\clearlist{location}
	\clearfield{volume}
	\clearfield{series}
	\clearfield{eprint}
	\clearfield{eprinttype}
 }{}
 }
\def\BibTeX{{\rm B\kern-.05em{\sc i\kern-.025em b}\kern-.08em
    T\kern-.1667em\lower.7ex\hbox{E}\kern-.125emX}}

\begin{document}
\title{Temporally Consistent Graph Q-Networks for Intelligent Network Control}

\author{
  Zacharias Veiksaar$^{1}$,
  Maxime Bouton$^{2}$\\
  \texttt{zacharias@veiksaar.se, maxime.bouton@ericsson.com}
  \thanks{
  $^{1}$ Work done while at Ericsson Research, Sweden. $^{2}$ Ericsson Research, Sweden.
  }
}

\maketitle
\thispagestyle{firstpage}
\begin{abstract}
Mobile networks continue to grow in complexity and next generation networks are expected to support both increasing traffic loads and more diverse services. As network complexity rises, optimizing antenna parameters under dynamic or changing objectives becomes increasingly challenging. We propose a novel multi-agent reinforcement learning (MARL) algorithm for high-level control and orchestration of mobile networks.
The Temporally Consistent Graph Q-Network (TC-GQN) algorithm learns a self-predicting representation of the whole network that is task-independent and aggregates information from all base-stations. A graph neural network is trained using a global reward function to assign coordinated local actions based on the learned encoding of the global network state.  
We evaluate the algorithm in a simulated environment to orchestrate an energy-saving feature across multiple sectors and multiple carriers under different quality of service (QoS) constraints. 
The proposed algorithm outperforms state-of-the-art graph-based baselines and a competitive rule-based controller by improving hardware sleep time while maintaining QoS. Moreover, the learned representation enables rapid adaptation to changing intents.

\end{abstract}


\section{Introduction}

Radio access networks are increasing in density and 6G is expected to support an increase in traffic as well as new services related to robotics, healthcare, and cloud gaming with strict throughput and latency constraints. To achieve this, the industry is shifting towards a new paradigm of intent-driven AI-native autonomous networks. Operators are expected to steer the network through high-level intents, and AI is seen as an enabling technology to realize this intent through real-time decision-making by changing network parameters~\cite{khani2024,ericsson2025intent}.

Optimizing the network based on high-level intent requires having coordinated policies for all network nodes as changing parameters in one node can greatly affect other parts of the network, for example by causing congestion in neighboring sites. In addition, network performance is highly dependent on traffic demand as well as radio propagation conditions which are hard to model. Therefore, heuristic strategies to realize intent-driven automation are 
often suboptimal and instead AI-based methods such as multi-agent reinforcement learning (MARL) have been proposed~\cite{bouton2023}. 

Prior work has applied MARL to coverage and capacity optimization using graph neural networks (GNNs) to coordinate antenna tilt control across network nodes~\cite{bouton2023} as well as to the domain of energy-saving~\cite{cai2024}.
However, these RL agents are typically task-specific, and as operators are expected to change intents dynamically, agents must adapt their policies accordingly. Existing algorithms capture intents via a reward function and thus require retraining whenever the intent changes. This incurs a significant cost in computation and data collection. 

Conversely, recent progress in single-agent RL has introduced self-supervised learning components to improve sample efficiency by learning an efficient representation of the environment. These methods have been highly successful on benchmark problems, improving performance and reducing data required for training~\cite{zhao2023, hansen2024}. However, integrating these methods with graph-based MARL to learn a transferable network representation that supports adaptation to new intents remains unexplored.

To bridge this gap, we propose TC-GQN, a novel algorithm that extends graph-based RL with self-supervised temporal consistency learning.
We pool graph-aware local observation embeddings from each agent into a global state using an attention mechanism, and encode this into a self-predictive latent state representation. By utilizing an auxiliary loss to predict inherent environment variables, rather than reward, TC-GQN decouples the prediction task from the reward function and yields a task-independent encoding of the environment dynamics. 
This allows us to maintain the coordination benefits of graph-based techniques while introducing the rapid adaptation capability of representation learning in a multi-agent setting.

In this work, we model intelligent network control as a cooperative MARL problem where intents on energy-saving and QoS are captured in a shared reward function, and agents can enable or disable carrier-sleep. In simulation, we show that the algorithm outperforms a rule-based baseline and graph-based MARL methods. We further test adaptation to new intents by adding more QoS constraints and comparing a pretrained agent with a randomly initialized agent, and find that the agent requires substantially less data and outperforms the baselines from the start when pretrained. The learned global network representation also changes little during fine-tuning which indicates that it captures network dynamics and remains informative across different intents.

\section{Related Work}
Reinforcement learning techniques have shown promising results when applied to various network control tasks. In the energy efficiency domain, deep RL techniques have successfully been applied to reduce energy consumption in networks under QoS constraints, both in single-agent settings \cite{ye2020} and in large MARL settings \cite{cai2024}. 
Similarly, graph-based neural networks have also been shown effective in coordinating multi-agent antenna tilt control, both through neighbor-enhanced observations \cite{jin2022} and by decomposing the importance of a global reward signal across nodes through value decomposition \cite{bouton2023}. However, these approaches are typically limited to optimizing a single predefined objective defined through a reward function, leaving them unable to adapt to changing network intents without retraining. 

To deploy such solutions, one must first address the fundamental scalability challenge of multi-agent environments. The dimensionality of the state and action spaces can grow quickly and make naïve approaches such as concatenating agent observations and actions become impractical. Such solutions also suffer from a lack of adaptability, for example, when the number of agents in the environment changes. Graph-based methods address this issue by enhancing single-agent algorithms with neighboring observations \cite{jin2022} or through more general coordinated algorithms that are robust to network topologies \cite{bouton2023}. Apart from scalability, another problem that arises in a multi-agent setting is the lack of individual reward signals that promote cooperation among agents. One method for proper reward assignment is to learn a decomposition of a global reward signal \cite{sunehag2017, rashid2018}, which can be combined with graph-based networks for cooperation.

However, relying solely on a scalar reward signal to inform complex cooperation behavior can be insufficient for learning a desired policy. To address this, representation learning in RL provides informative latent states by training an encoder with an auxiliary task. These tasks can be predicting future states, creating representations that capture the transition dynamics of an environment, which improves sample efficiency and generalization. This latent space can be used either to augment the observations of a traditional RL algorithm~\cite{zhao2023,schwarzer2021} or for model predictive control \cite{hansen2024} and has been highly effective in such single-agent RL settings.

Our work bridges the gap between these domains by introducing TC-GQN, an algorithm which extends graph-based MARL with a self-supervised component that learns a task-agnostic representation of the global network state. By training the network to predict inherent environment variables rather than rewards, we decouple the learned representation from the immediate task and instead focus on encoding the underlying environment dynamics. This allows for rapid adaptation to new intents while maintaining the coordination  benefits of graph-based techniques.

\section{Background} 
\subsection{Multi-Agent Reinforcement Learning}

Cooperative MARL can be defined as a decentralized partially observable Markov decision process (dec-POMDP). 
A dec-POMDP is formally defined by the tuple $(n, \mathcal{S}, \{\mathcal{A}_i\}_{i=1}^{n}, \{\mathcal{O}_i\}_{i=1}^{n}, P, O, R, \gamma)$, where $n$ is the number of agents, $\mathcal{S}$ is the global state space, $\mathcal{A}_i$ is the action space of agent $i$, and $\mathcal{O}_i$ is the observation space of agent $i$. 
$P(s_{t+1}\mid s_t,a_t^1,\ldots,a_t^n)$ is the unknown transition probability of the global state given the previous state and the agents' actions, and $O(o_{t+1}^1, \ldots, o_{t+1}^n \mid s_{t+1}, a_t^1, \ldots, a^n_t)$ describes the unknown probability of each agent receiving a local observation given the global state and actions.  The shared reward function $R$ outputs a reward at each step $r_t = R(s_{t+1}, a_t^1, \ldots, a_t^n) \in \mathbb{R}$ and $\gamma \in [0,1)$ is a discount factor for future rewards. 

The goal of our algorithm is to find a joint policy $\pi: \mathcal{O}_1 \times \ldots \times \mathcal{O}_n \rightarrow \mathcal{A}_1\times \ldots \times \mathcal{A}_n$ that maps local observations to actions and maximizes  $E[\sum_{t=0}^{\infty} \gamma^t r_{t+1} \mid s_0]$ where $r_t$ is the shared reward obtained by following the policy. Often, a value function $Q$ can be learned to estimate this quantity, and a policy can then be derived by maximizing the Q-function.

\subsection{Deep Learning on Sets}

Agents can be represented by an undirected graph or a set, and an adaptable algorithm must be invariant to the ordering of agent observations. To represent the Q-function as a parametric function, deep learning architectures such as attention mechanisms and graph neural networks are particularly useful~\cite{lee2019}. We use GNNs to transform the state of the environment into local embedding vectors which contain information about neighboring agents according to a graph topology provided by the environment. In particular, we adopt the graph attention network architecture \cite{velickovic2018}, which uses self-attention to identify the most relevant neighbor observations and supports a variable number of neighbors.
Since GNNs output graphs with vector embeddings, we compress them into a global latent network state using an attention-based pooling mechanism~\cite{lee2019} detailed in \cref{sec:algorithm}.

\section{Intelligent RAN Energy Control as MARL}\label{sec:problem}

\begin{figure}
    \centering
    \includegraphics[width=0.8\columnwidth]{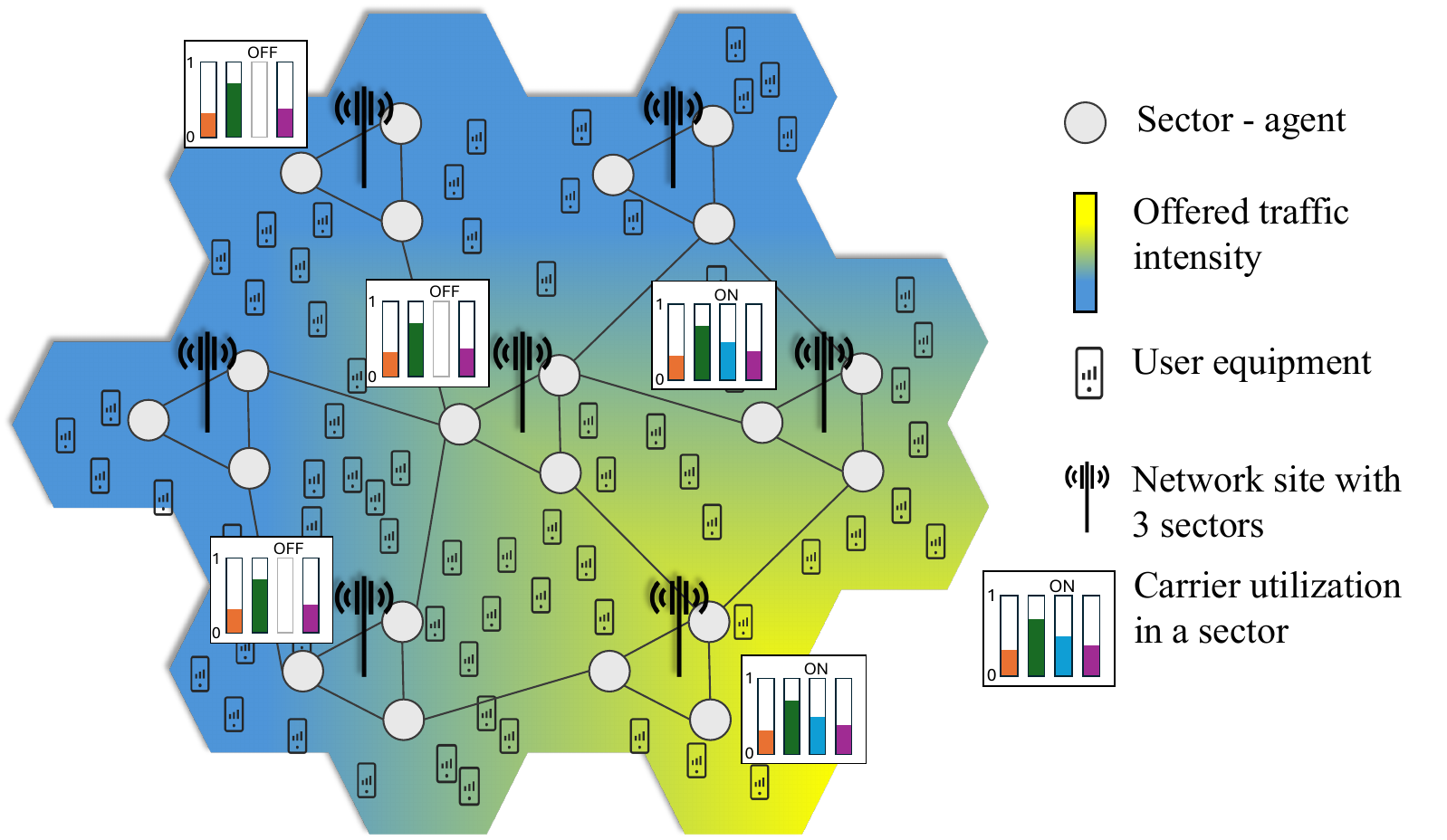}
    \caption{Illustration of the RAN with 21 sectors and 4 carriers per sector and an associated graph. Traffic is distributed non-uniformly over the map.}
    \label{fig:network}
\end{figure}

\subsection{System Model}

To evaluate the benefit of MARL with global network representation learning for RAN control, we  consider an energy-saving problem in which certain carriers in the network can be shut down to reduce energy consumption during periods of low traffic demand. Disabling a carrier may offload users to neighboring cells, so decisions must account for the entire network rather than a single cell. Energy-savings must be balanced with maintaining stringent service requirements and this trade-off is expected to be changed dynamically based on operator intent, requiring the AI agent to adapt quickly.

The RAN consists of multiple geographical sectors equipped with multiple base-station antennas operating at different frequency bands, and user equipment connects through one or more of these frequencies. A given frequency in a specific sector, and its connected users, is referred to as a cell.

We consider a geographical area with a hexagonal deployment consisting of 21 sectors as illustrated in \cref{fig:network}. Each sector is equipped with four carriers (also known as layers): \SI{800}{\mega\hertz}, \SI{1800}{\mega\hertz}, \SI{2100}{\mega\hertz}, and \SI{3500}{\mega\hertz} corresponding to one low- and mid-band coverage frequency, and two mid-band capacity layers referred to as L08, L18, L21, and L35 respectively. They have bandwidths of \SI{10}{\mega\hertz}, \SI{20}{\mega\hertz}, \SI{20}{\mega\hertz}, and \SI{100}{\mega\hertz} respectively.

The L21 carrier is assumed to be equipped with an energy-saving feature running on the base-station that can be activated remotely. In a real deployment, when activated, the feature could intermittently turn the cell on or off as fast as every millisecond~\cite{cai2024}. For simplicity, we model the feature as keeping the cell off all the time when it is activated. We assume that our proposed algorithm is implemented as a non-real time radio intelligent controller in an open-RAN architecture~\cite{khani2024}. Its role is to orchestrate the energy-saving features to satisfy high-level intents such as maintaining a good quality of service. This centralized controller needs to receive performance counters from each network node and typically operate at a frequency ranging from minutes to days. We assume that this centralized controller takes a decision every hour to enable or disable a local energy-saving feature. 

A total of \num{1000} users are placed uniformly on the map and randomly labeled indoor or outdoor to yield 80\% indoor traffic. Users generate downlink traffic according to a Weibull distribution creating hotspots in the map \cite{lee2014}. The intersite distance is sampled uniformly between \SI{400}{\meter} and \SI{1000}{\meter} at the start of each training or evaluation episode (see \cref{sec:experiments}). To model temporal traffic variation, we sample an hourly total traffic demand over the area from a dataset collected across more than \num{100} live 5G base stations.

Users connect to the carrier providing the highest achievable bitrate. Bitrate is estimated using 3GPP-compliant models from the received signal to noise ratio per carrier. Signal strength is computed using the site specific propagation model of \citeauthor{asplund2018}~\cite{asplund2018}. The number of connected users per cell, the offered traffic per user, and their estimated throughput are then used to compute key performance indicators (KPIs).

Our system model relies on three main KPIs: carrier utilization, throughput, and power consumption.
The utilization of a carrier $c$ associated with agent $i$ is $U^i_c$ and corresponds
$\text{U}^i_c = \frac{\sum_u n_u}{N_{\text{PRB}c}}$
where $n_u$ is the number of PRBs associated with user $u$ and depends on the user throughput $\rho_u$, and $N_{\text{PRB}c}$ is the total number of PRBs available at the carrier, determined by its bandwidth.
Throughput is defined per carrier and corresponds to the average throughput of the users connected to that carrier: 
$\tau^i_c = \frac{1}{|\mathbb{U}^i_c|} \sum_{u\in\mathbb{U}^i_c} \rho_u$
where $\mathbb{U}^i_c$ represents the set of users connect to carrier $c$ in sector $i$.
Finally, to estimate the power consumption of each RAN sector, we adopt a model in which the total power is the sum of the idle power consumption of the radio at each carrier and a load-dependent component scaling approximately linearly with the utilization, as recommended in ETSI TR 132 972 \cite{ETSI_TR_132972_V18}.

\subsection{MARL Formulation}

To formulate the network control problem as MARL we need to define the agents, the observation spaces and action spaces of each agent, and the shared reward function. 
The objective is to turn antennas off to save energy while maintaining a minimum level of throughput in a given network area with multiple base-stations. This creates a complex MARL problem where agents must coordinate to minimize energy consumption without disrupting user connectivity. We use the energy-saving problem described in the previous section as an example, with the observations, actions, and rewards being specific design choices. However, we anticipate that the proposed algorithm will generalize to other variants of this network control problem, such as tilt optimization~\cite{bouton2023}.

We model the problem as a discrete task where each agent selects an action at each time step. During a step $t$, our algorithm observes the network topology, and gets local observation vectors, $o^1_t, ..., o^{n}_t$. It performs inference and returns actions for all sectors, $a^1_t, ... a^{n}_t$ and observes the next observations and the shared rewards which are added to a replay buffer as described in the next section. 

We define agents as sectors in the network. In our test hexagonal deployment there are thus 21 agents. Each agent is equipped with the four carriers described above.

The observation space corresponds to network KPIs for each carrier. Each agent observes the utilization of each carrier (four values), the average user throughput in each carrier (four values), the state of the energy-saving feature taking values -1 and 1. Finally, it also observes an encoding of the hour of the day: $[\sin(\frac{2\pi h}{24}),\cos(\frac{2\pi h}{24})]$ where $h\in \{0,1,\dots, 23\}$ is the hour of the day. The observation space of each agent is thus an 11-dimensional vector. In addition, our algorithm observes the graph topology $\Gamma_t$. The topology is determined via geographical proximity as well as orientation of each sector. Two sites are connected if they are less than \SI{2}{\kilo\meter} from each other via the sectors facing each other, and each co-site sector is connected, as illustrated in \cref{fig:network}. In a live network, the graph could be obtained through standard feature like automatic neighbor relations based on the hand-over between cells. Our algorithm is agnostic to the specific graph size and topology.

The action space is discrete and has only two values: ON and OFF. It corresponds to activating the energy feature on the L21 carrier. When the OFF action is selected we say that the carrier is considered to be in a sleep state.

The shared reward is defined as a sum of local reward components. We define $R^i(s_{t+1}, o^i_t, a^i_t)$ as the local reward associated to agent $i$. Note that this local reward depends on the unobserved global state which can be affected by the actions of other agents through possible offloading of users between sectors. $R^i$ is defined using indicator functions as follows: 
$[\mathbf{1}_{[\tau_{i,L18}>50\text{Mbps}]} - \lambda\mathbf{1}_{[\tau_{i,L18}\leq50\text{Mbps}]} ]\mathbf{1}_{[a_i =\text{off}]} - \tilde{P_i}\mathbf{1}_{[a_i =\text{on}]}$
where $\mathbf{1}$ is the indicator function, $\tau_{i,L18}$ represents the average user throughput in the L18 carrier for agent $i$, and $\tilde{P_i}$ is the power consumption of the whole sector (considering all 4 carriers) normalized to be between 0 and 1. 
In simple terms, the reward gives a bonus of 1 when the agent turns on sleep and good throughput is achieved and gives a penalty of $-\lambda$ when the agent enables sleep and causes poor throughput. When the agent does not turn on sleep, the reward is proportional to the power consumption. 
The total shared reward is the sum of the individual rewards: $R=\sum_{i=1}^n R^i$.

In an intent-based autonomous network, this reward function may change based on high-level intents. For example, the 50Mbps constraint on throughput may be changed by the operator, it may be defined on a different carrier than the L18 one, or simply the $\lambda$ parameter may be adjusted. We expect these changes to lead to different behavior when we solve the MARL problem. With standard algorithms, however, such changes in intent require retraining the agent from scratch.

\section{Proposed Algorithm}
\label{sec:algorithm}

\begin{figure}[t]
    \centering
    \includegraphics[width=0.8\columnwidth]{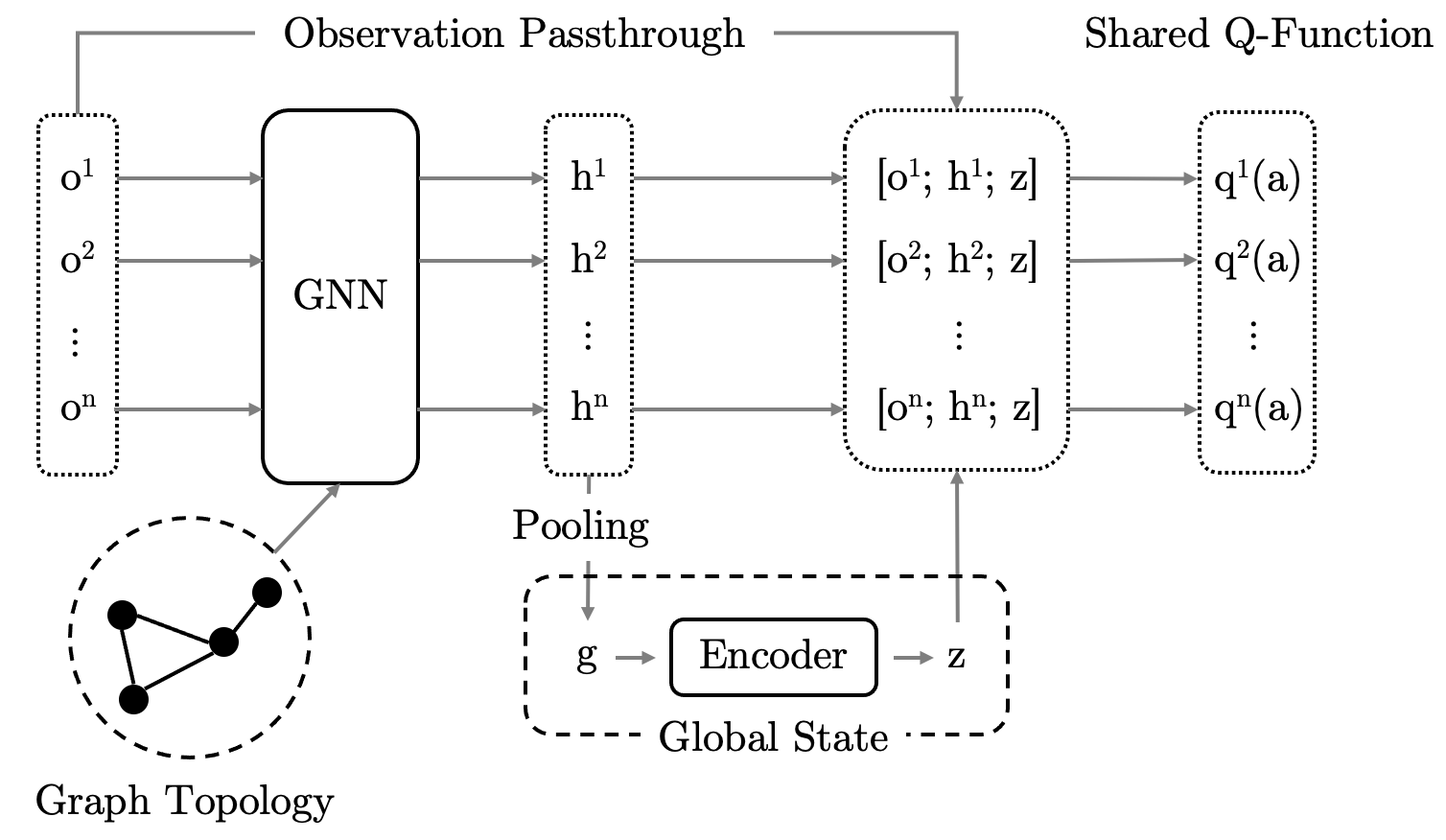}
    \caption{Diagram of the proposed algorithm during inference.}
    \label{fig:inference_diagram}
\end{figure}

\subsection{Components and Training}
The proposed algorithm uses a graph-based Q-network architecture that is augmented to ingest a global state representation learned through the prediction-based auxiliary task, as illustrated in \cref{fig:inference_diagram}. To generate this representation, we first pool agent-wise information using an attention mechanism to output a global graph state. This state is then passed through an encoder to generate a temporally consistent global state representation. 

Although we validate the TC-GQN algorithm using network control tasks, the architecture generalizes to any graph-based MARL problem where a graph topology is available.

The main components of the algorithm are:
\begin{align*}
     \text{GNN} &: \{h_t^k\}_{k=1}^n = G_\theta \bigl(\{o_t^k\}_{k=1}^n, \Gamma_t\bigr)\\
    \text{State Pooling} &: g_t = A^{state}_\phi\bigl(\{h_t^k\}_{k=1}^n\bigr)\\
    \text{Action Pooling} &: u_t = A^{action}_\chi\bigl(\{a_t^k\}_{k=1}^n\bigr)\\
    \text{Encoder} &: z_t = E_\psi (g_t)\\
    \text{Dynamics} &: (z_{t+1}, p_{t+1}) = D_\eta(z_t, u_t)\\
    \text{Value} &: q_t^k(a)=Q_\omega(o_t^k, h_t^k, z_t)
\end{align*}
The GNN component first transforms agent observations using a simple MLP and then enriches this embedding with context from the observations of the agent's neighbors using a GNN. These encoded observations $h_t^k$ are then pooled over the agent dimension using attention based pooling to generate a global graph state representation $g_t$. It is this global graph representation $g_t$ that then acts as an input to the representation learning where we generate the temporally consistent global graph encoding $z_t$. To generate $q$-values for each agent we concatenate $o_t^k, h_t^k,$ and $z_t$ which together act as the input to a standard DQN architecture. We keep target networks for the GNN, pooling, encoder, and value components which are denoted with a superscript minus sign, e.g. $\theta^-$.

The proposed algorithm is an off-policy algorithm and experience is gathered through interaction with the environment using an epsilon-greedy exploration scheme. Gathered experience is stored in a replay buffer for access during the training pass where gradient updates are computed from a Q-function loss and an auxiliary loss, both of which are explained in more detail in the subsequent sections. After gradient updates are completed target networks are updated using Polyak averaging.

Introducing the self-supervised prediction objective increases training-time computational cost relative to the closest benchmark algorithm GQN \cite{bouton2023} due to the additional forward passes required for prediction and auxiliary loss evaluation. However, the complexity during inference is only marginally increased as the algorithm requires a single additional forward pass through the global encoder without any prediction taking place. This makes the algorithm close to parity with a standard GQN architecture in terms of inference latency.

\subsection{Learning the Encoder}

\begin{figure}[t]
    \centering
    \includegraphics[width=0.7\columnwidth]{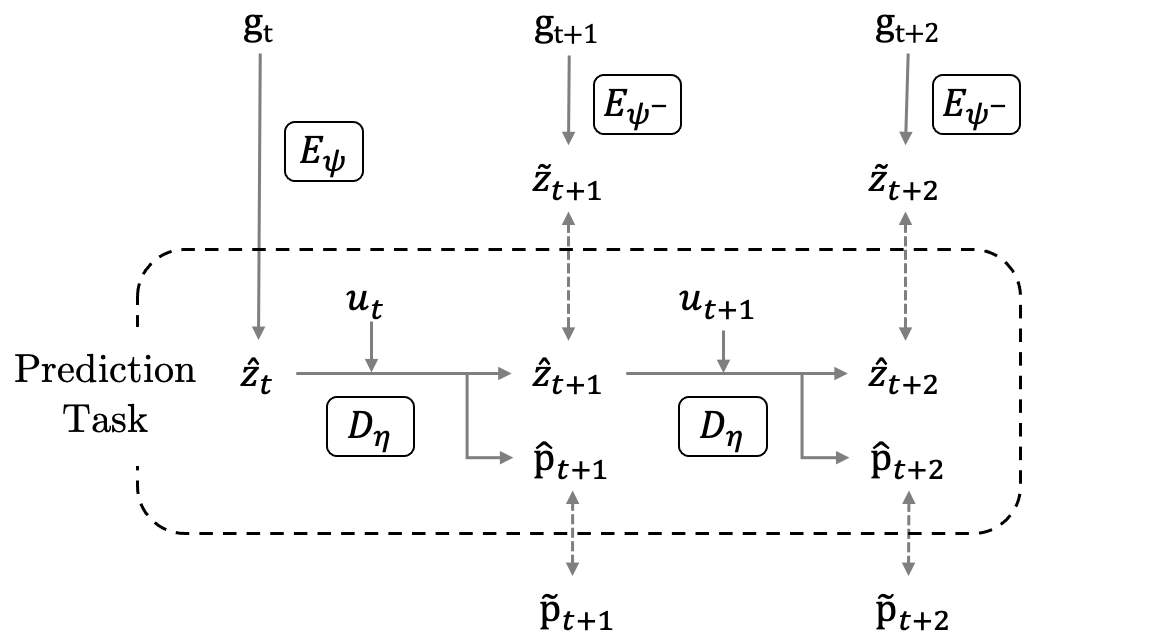}
    \caption{Diagram of the prediction task with a horizon of $H=2$ timesteps.}
    \label{fig:prediction_diagram}
\end{figure}

We separate the algorithm into two main parts, the encoder and the value function. The encoding process encompasses the GNN, state and action pooling, encoder, and dynamics components with the corresponding parameters $\theta, \phi, \chi,\psi,$ and $\eta$. To train the encoder we employ a combined state and target prediction which is described in \eqref{eq:state_reward_loss} and denoted by $\mathcal L_{aux}$.

We begin the training process by generating GNN embeddings using the agent observations $\{o_t^k\}_{k=1}^n$ and the global graph $\Gamma_t$ as $\{ h_t^k\}_{k=1}^n = G_{\theta}\bigl(\{o_t^k\}_{k=1}^n, \Gamma_t\bigr)$,
and pool these as
$g_t = A^{state}_{\phi}\bigl(\{ h_t^k\}_{k=1}^n\bigr)$.
The attention-based pooling mechanism is implemented by using a learnable query vector which effectively compresses the agent dimension to a single global state representation vector \cite{lee2019}.

To generate predicted future states $\hat z_{t+h}$ for $h=1,..., H$, where $H$ is the prediction horizon length, we begin by encoding the initial state $\hat z_t=E_\psi(g_t)$. To inform the dynamics model which will generate predictions we use learnable embeddings to turn agent actions into continuous vectors and then pool these as
$u_t = A^{action}_{\chi}\bigl(\{ a_t^k\}_{k=1}^n\bigr)$,
where the attention pooling for actions works analogously to the state pooling.
The predictions are then generated autoregressively, such that
$(\hat z_{t+h}, \hat p _{t+h}) =D_\eta(\hat z_{t+h-1}, u_{t+h-1})$
for $h \geq 1$. Here $\hat z_{t+h}$ and $\hat p _{t+h}$ are generated using separate MLPs with the same input. 
To generate latent state targets $\tilde z_t$ for these predictions we generate target GNN embeddings by
$\{\tilde h_t^k\}_{k=1}^n = G_{\theta^-}\bigl(\{o_t^k\}_{k=1}^n, \Gamma_t\bigr)$
which are then pooled as
$\tilde g_t = A^{state}_{\phi^-}\bigl(\{ \tilde h_t^k\}_{k=1}^n\bigr)$.
Using these target global graph embeddings we then compute the target embeddings $\tilde z_{t:t+H}$ for the interval $t:t+H$ by $\tilde z_t = E_{\psi^-}(\tilde g_t)$. The prediction target reference values $\tilde p_t$ are simply the actual target values given by the environment.
The learning of the encoder and dynamics components is illustrated in \cref{fig:prediction_diagram}.

Taken together we formulate the auxiliary loss $\mathcal L_{aux}$ as 
\begin{equation}\label{eq:state_reward_loss}
\sum_{h=1}^{H} \gamma_{aux}^{h} \left[ 
\left\| \hat{p}_{t+h} - \tilde{p}_{t+h} \right\|_2^{2}
-
\left( \frac{\hat{z}_{t+h}}{\left\|\hat{z}_{t+h}\right\|_2} \right)^{\top}
\left( \frac{\tilde{z}_{t+h}}{\left\|\tilde{z}_{t+h}\right\|_2} \right)
\right]
\end{equation}
where $\gamma_{aux}\in[0,1)$ is the discount rate. We use the cosine similarity loss for the state prediction task as it has been shown to achieve better performance with more stability compared to an MSE loss \cite{zhao2023}.

\begin{figure*}[!h]
    \centering
    \includegraphics[width=0.9\textwidth]{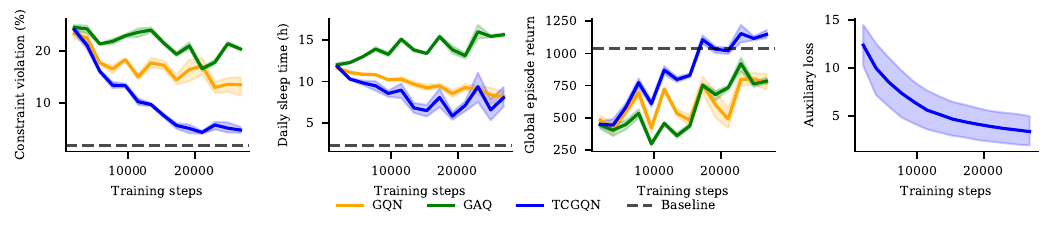}
    \caption{Performance of MARL agents during training in terms of sleep time (the higher the better) and probability of constraint violation (the lower the better).}
    \label{fig:benchmark}
\end{figure*}

\subsection{Learning the Value Function}
For learning the value function parameters $\omega$ we adopt a DQN architecture using double-Q learning with dueling, where the input consists of the raw agent observations, the enriched graph aware agent observation embeddings $h_t^k$, and the temporally consistent global graph representation $z_t$. We generate the Q-value function predictions individually for each agent by
$q_t^k(a^k_t)= Q_\omega\bigl(\operatorname{sg}(o_t^k), \operatorname{sg}(h_t^k), \operatorname{sg}(z_t)\bigr)$,
where $\operatorname{sg}(\cdot)$ is the stop-gradient operator. We then pool these Q-value functions by taking the sum over all agents to generate the group level reward prediction $\hat q_t(\textbf{a})=\sum_{k=1}^nq_t^k(a^k)$ as in value decomposition networks~\cite{sunehag2017}. The target Q-values $\tilde q_t(a)$ are generated by first finding the online networks prediction for the best action as 
$\textbf{a}^* = \operatorname{argmax}_{\textbf{a}\in \mathcal A} \hat q_{t+1}(\textbf{a})$, where $\textbf{a}$ is the vector of agent actions.
We find the target Q-values through
$\tilde q_t^k(a_t)= Q_{\omega^-}\bigl(\operatorname{sg}(o_t^k), \operatorname{sg}(h_t^k), \operatorname{sg}(z_t)\bigr)$
where we use the $h_t^k$ and $z_t$ generated by the online encoding networks. We then define target Q-value function by $\hat q_t^-(\textbf{a})=\sum_{k=1}^n \tilde q_t^k(a^k)$ and let
$\tilde q_t = r_t+ \gamma_q \cdot \hat q_{t+1}^-(\textbf{a}^*)$
where $r_t$ is the actual group level reward from the environment and $\gamma_{q}\in[0,1)$ is the discount rate. The Q-loss is then given by
$\mathcal L_Q= \bigl(\hat q_t(\textbf{a}_t) - \tilde q_t\bigr)^2$.
An important distinction here is that the training of the two sets of parameters, for the encoding and the Q-function respectively, is separated and each trained by its own optimizer. We also employ the stop-gradient operator before computing the Q-values so as to completely separate the two training processes.

\section{Experiments}\label{sec:experiments}

\subsection{Experiment Setup}

We implement the MARL network control environment in a proprietary simulator environment following the system model described in \cref{sec:problem}. 
The simulator is used to generate synthetic episodes corresponding to a specific intersite distance sampled uniformly, a realization of the Weibull distribution modeling the traffic distribution in space, a realization of indoor/outdoor traffic distribution, and a specific temporal traffic pattern. 
The episode consists of 96 steps where each step spans one hour of network utilization.

We compare different intelligent network control algorithms according to three different metrics. The first is the \emph{constraint violation probability} which is the number of steps where the throughput constraint is not satisfied divided by the total number of steps in an episode. The second metric is the \emph{average daily sleep time} per sector which is the number of steps where the sleep action was taken without violating the constraint divided by the number of days and sector. The final metric is the \emph{return} which is the shared reward summed over all episode steps.

For TC-GQN, we set the prediction targets $p_t$ to be the mean throughput per agent for L08, L18, and L35, and set hyperparameters as listed in \cref{tab:hyperparams}.

\begin{table}[h]
    \centering
    \caption{Hyperparameters}
    \label{tab:hyperparams}
    \begin{small}
    \begin{tabular}{lr}
        \toprule
        \textbf{Parameter} & \textbf{Value} \\
        \midrule
        \textit{Training \& Optimization} & \\
        \quad Total Timesteps & $25\,000$ \\
        \quad Learning Rate (Adam) & $10^{-4}$ \\
        \quad Batch Size & 256 \\
        \quad Discount Factor ($\gamma$) & 0.95 \\
        \quad Polyak Coeff. ($\tau$) & 0.005 \\
        \quad Gradient Clip & 40 \\
        \quad Exploration ($\epsilon$) & Linear($1.0,\ 0.01,\ 17\  500$) \\
        \quad Parallel Workers & 20 \\
        \quad Min. Sampling Steps & $1\,000$ \\
        \quad Replay Buffer & Unlimited \\
        \midrule
        \textit{Architecture Specifics} & \\
        \quad GNN Type & GATv2Conv (4 heads) \\
        \quad Embedding Dim. & 10 \\
        \quad Prediction Horizon & 5 \\
        \midrule
        \textit{Hidden Layer Sizes} & \\
        \quad GNN / Enc. / Dyn. / Q-Head & $[64, 64]$ \\
        \quad Pre-GNN / Dueling Heads & $[64]$ \\
        \quad Activations & ReLU \\
        \bottomrule
    \end{tabular}
    \end{small}
\end{table}

\subsection{Performance of TC-GQN for RAN Energy-Saving}\label{section:benchmarking}

To evaluate our proposed TC-GQN algorithm, we conduct a comparative study on the energy-saving task against three other methods: (1) a heuristic baseline controller that takes OFF actions at night time and if the capacity cell utilization is below 10\%, and takes ON actions if the utilization of the mid-band coverage cell is above 25\%, (2) the GAQ algorithm which incorporates neighboring agent observations but which does not have a shared reward signal and uses the local reward $R^i$ instead \cite{jin2022}, and (3) the GQN algorithm that incorporates all other agent observations and employs a shared reward \cite{bouton2023}.

Each algorithm is trained on \num{25000} steps with three random seeds and we report the mean and 95\% confidence interval. The baseline is evaluated on \num{20} episodes. The results of the experiment are presented in \cref{fig:benchmark} and we note that the auxiliary loss is well behaved and steadily decreases during the experiment, indicating that adding the temporal consistency objective does not destabilize training. We also note that TC-GQN achieves high episode returns faster and ends up outperforming all of the competing methods.

Looking at the two network performance metrics \emph{constraint violation probability} and \emph{sleep time}, the heuristic provides a conservative approach, while the RL algorithms converge to more aggressive strategies. Notably both GQN and TC-GQN converge to a strategy which achieves roughly eight hours of daily sleep time, but where TC-GQN violates the constraint only about a third as often as GQN. As these two algorithms are the most similar in the benchmark it highlights the benefit of introducing temporal consistency.

While GAQ and GQN converge to drastically different strategies, the two algorithms achieve the same episode reward in the end. This suggests that the reward signal might not clearly favor a certain type of behavior, potentially making it more difficult for an agent to learn an optimal policy. Despite this apparent ambiguity, however, TC-GQN is still able to learn a policy that performs well across all metrics.

\subsection{Adapting to Different Intents}

\begin{figure*}[!t]
    \centering
    \includegraphics[width=0.9\textwidth]{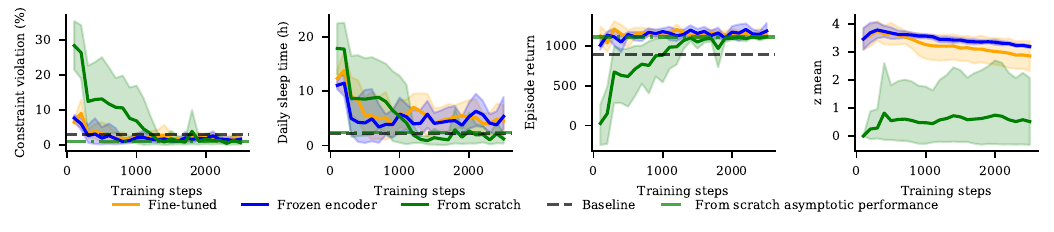}
    \caption{Performance of TCGQN fine tuned and from scratch when transferred on a new task with additional constraints on all the carriers.}
    \label{fig:transfer_learning}
\end{figure*}

A key advantage of training the temporal consistency encoder independently of the reward signal is that the learned representation can be reused to quickly adapt to new adjacent tasks. Since the encoder is trained to predict network dynamics via the carrier throughputs at the next step rather than task-specific rewards, the learned representation should remain robust to changing reward functions, provided that the prediction target still has relevance for the new task. 

To highlight this capability we conduct an adaptation experiment to compare how fast an agent pretrained on the energy-saving task (Section \ref{section:benchmarking}) can adapt to satisfy more stringent QoS constraints. We compare three different training strategies: (1) From scratch, where a TC-GQN agent is initialized with random weights, (2) Fine-Tuning, where an agent is initialized with pretrained weights and all weights are trainable, and (3) Frozen Encoder, a pretrained agent where only the value function weights are trainable while the encoder is frozen. We set the number of training steps to 
\num{2500} which is 10 times less data than required for full training.

The environment is changed to add two new throughput constraints on the L08 and L35 carriers. It is achieved by adding the following terms in $R^i$: $-\lambda\mathbf{1}_{[\tau_{i,L08}\leq15\text{Mbps}]} \mathbf{1}_{[a_i =\text{off}]}$ and $-\lambda\mathbf{1}_{[\tau_{i,L35}\leq100\text{Mbps}]} \mathbf{1}_{[a_i =\text{off}]}$ which penalizes the agent for causing the throughput to be below 15 Mbps for the low frequency carrier and 100 Mbps for the high frequency high bandwidth carrier. These throughput threshold values are used to model operator intents for quality of service. They are expected to change when new services are being rolled out. 

The results of the experiment are presented in \cref{fig:transfer_learning} where the pretrained agents clearly converge much faster than the agent training from scratch, suggesting that the learned global state representation is transferrable across certain changes of intent. We also note that the mean value of the global state representation $z$ remains similar between the two pretrained agents compared to the one learning from scratch, further indicating that the pretrained global state representation already contained elements that generalize across tasks.

\section{Conclusions}
We propose Temporally Consistent Graph Q-Networks (TC-GQN), a graph-based MARL algorithm that learns a task-agnostic global network representation using self-supervised representation learning. On a simulated RAN energy-saving task, TC-GQN outperforms other graph-based RL methods and a rule-based method.  It creates the most cell sleep opportunity with less violation of the QoS constraint.
In adaptation experiments where we change the intent by making QoS constraints more stringent, a pretrained TC-GQN outperforms the baseline from the start and requires less training to reach a strategy fitting the new goal. We also find that the learned global network representation remains relatively unchanged during fine-tuning. These results suggest that the learned encoder captures network dynamics and can transfer to adjacent tasks with different intents, paving the way towards AI-native intent-driven network automation. In today's networks, this algorithm could be adopted in the non real-time RIC of an O-RAN network and use existing O1 and A1 interfaces to observe cell performance and send actions \cite{khani2024}. A standard MLOps framework could be responsible for triggering fine-tuning upon changes in service requirements intents.

Several limitations and areas of future work need to be addressed to fully adopt this technology in 6G networks. The encoder learns to predict future states solely from the current one, with no access to recent observation history. Incorporating short-term temporal context may allow for more accurate predictions of future states and thus improve the performance of the algorithm, especially for network control tasks requiring good KPI forecasting capabilities. Finally, we observe an apparent robustness of TC-GQN to reward misalignment with the network optimization intent, however, more systematic reward design from intent should be further investigated.

\printbibliography
\end{document}